\documentclass[twocolumn,showpacs,preprintnumbers,amsmath,amssymb]{revtex4}
\usepackage{graphicx}
\usepackage{dcolumn}
\usepackage{bm}

\begin{document}
\def\etal{{\it et al.}}
\def\go{\rightarrow  }
\def\be{\begin{equation}}
\def\ee{\end{equation}}
\def\br{\begin{eqnarray}}
\def\er{\end{eqnarray}}
\def\brn{\begin{eqnarray*}}
\def\ern{\end{eqnarray*}}
\def\rf#1{{(\ref{#1})}}
\def\a {{\alpha}}
\def\b {{\beta}}
\def\e {{\epsilon}}
\def\k {{\kappa}}
\def\s {{\sigma}}
\def\w {{\omega}}
\def\sss{\scriptscriptstyle}
\def\nn{\nonumber}
\def\ie{{\em i.e., }}
\def\x{\times}
\def\F {{{\cal F}}}
\def\M {{{\cal M}}}
\def\T {{{\cal T}}}
\def\pb {{\bf p}}
\def\qb {{\bf q}}
\def\kb {{\bf k}}


\preprint{}

\title{Detection of supernovae neutrinos
with neutrino-iron scattering}

\author{A.R. Samana}
\author{C.A. Bertulani}%
\affiliation{%
Department of Physics, Texas A\&M University Commerce,\\
P.O.3011 Commerce, 75429 TX, USA
}%

\date{\today}

\begin{abstract}
The  $\nu_e-^{56}$Fe cross section  is evaluated
in the projected quasiparticle random phase approximation (PQRPA).
This model  solves the puzzle observed in RPA for nuclei with mass
around $^{12}$C, because it is the only RPA model that
treats the Pauli principle correctly. The cross sections
as a function of the incident neutrino energy are compared with
recent theoretical calculations of similar models. The average
cross section weighted with the flux spectrum yields a good
agreement with the experimental data.
The expected number of events in the detection of supernova neutrinos
is calculated for the LVD detector
leading to an upper limit for the electron neutrino energy
of particular importance in this experiment.
\end{abstract}


\maketitle

A careful knowledge of the semileptonic weak interactions in nuclei
allows the possibility of testing implications of physics beyond the
standard model, such as exotic properties of neutrino oscillations
and massiveness. The dynamics of supernova collapse and explosions
as well as the synthesis of heavy nuclei are strongly dominated by
neutrinos.
For example, neutrinos carry away about 99\% of gravitational binding
energy in the core-collapse of a massive star, and only a
small fraction  ($\sim 1\%$) is transferred to the stalled shock
front, creating ejected neutrino fluxes observed in supernova
remnants~\cite{Dun06}.

It was shown in Ref.~\cite{Sam06} that accurate nuclear structure
calculations are essential to constrain the neutrino oscillations
parameters of the LSND experiment \cite{Ath98}. This  was also noted
in previous works, e.g. Hayes~\etal~  in Ref.~\cite{Hay00}. In that
work, based on a shell model, the same exclusive cross section in
$^{12}$C is obtained as with another shell model used in
Ref~\cite{Vol00}. This shows the importance of including
configuration mixing (as done in both references) for this nucleus
\footnote{The importance of configuration mixing in $^{12}$C is
known since the very first work in $p$-shell nuclei by Cohen and
Kurath in '65~\cite{Coh65}.}. Nevertheless the QRPA predictions of
Ref.~\cite{Vol00} do not yield good results for this nucleus because
the configuration mixing is not properly accounted for and the
projection procedure (as done in Ref.~\cite{Krm05}) is not included.
In particular, the employment of PQRPA for the inclusive
$^{12}$C$(\nu_e,e^-)^{12}$N cross section, instead of the continuum
RPA (CRPA) used by the LSND collaboration in the analysis of
${\nu}_\mu \go{\nu}_e$ oscillations of the 1993-1995 data sample,
leads to an increased oscillation probability. Then, the previously
found consistency between the $(\sin^2 2\theta, \Delta m^2)$
confidence level regions for the ${\nu}_\mu\go {\nu}_e$ and the
$\bar{\nu}_\mu\go \bar{\nu}_e$ oscillations is
decreased~\cite{Sam06}.

The measured observables are flux-averaged cross sections.
The KARMEN Collaboration measured
charged and neutral cross sections induced on $^{12}$C~\cite{Mas98}.
They also measured (the only experimental data
for a medium-heavy nucleus)
the neutrino reaction $^{56}$Fe$(\nu_e,e^-)^{56}$Co from
e$^-$-bremsstrahlung
with the detector surrounding shield~\cite{Mas99}.
This cross section is important to test the ability of
nuclear models in explaining reactions on nuclei with
masses around iron, which play an important role
in supernova collapse~\cite{Woo90}.
Experiments on neutrino oscillations
such as MINOS~\cite{Ada07}
use iron as material detector, and future experiments,
such as SNS at ORNL \cite{Efr05} plan to use
the same material.

In a recent work, Agafonova \etal~\cite{Aga07} studied the effect of
neutrino oscillations on the supernova  neutrino signal with the LVD
detector.
This detector studies supernova neutrinos through their interactions with
protons and carbon nuclei in a liquid scintillator and with iron nuclei
in the support structure.
Several estimates on
deviations of the detected signal arising from different
constraints on the astrophysical parameters, oscillation parameters
and the non-thermal nature of the neutrino fluxes were studied before~\cite{Aga07}.
Nevertheless, in all
their estimates the corresponding $\nu$-nucleus cross sections
were kept within strict limits.

In this work, we calculate the $\nu_e-^{56}$Fe cross sections
using QRPA and PQRPA models to
account for allowed and forbidden transitions.
The present calculations are the first within the PQRPA framework
for this purpose.
In Ref.~\cite{Krm02} it was established that PQRPA is
the proper theory to treat both short range  pairing and
long range random-phase (RPA) correlations.
When QRPA was implemented for the triad
$\{{{^{12}{\rm B}},{^{12}{\rm C}},{^{12}{\rm N}}}\}$
there were difficulties in choosing the ground state of $^{12}$N because
the lowest state is not the most collective~\cite{Vol00}. PQRPA
solves this puzzle because it treats correctly
the Pauli Principle, yielding better results for the distribution of the
Gamow-Teller (GT) strength.
This problem does not exist in heavier nuclei,
where the neutron excess allows QRPA to
account for pairing and RPA correlations~\cite{Krm93}.
In the case of medium-heavy nuclei,
such as $^{56}$Fe, the consequences of the projection technique
procedure can be manifest.

Many calculations of the $^{56}$Fe$(\nu_e,e^-)^{56}$Co cross sections
with microscopic and global models have been reported previously.
The first were shell model calculations developed by
Bugaev \etal~\cite{Bug79}. They
obtained the $\nu$-nucleus cross sections as a function of the
incident neutrino energy. A second estimate was obtained by
Kolbe \etal~\cite{Kol99a} using a nuclear Hybrid model: shell model
for the GT and Fermi (F) transitions,  and continuum  RPA
(CRPA) for forbidden transitions. This cross section was employed
to estimate the number of events from $\nu$-$^{56}$Fe reactions
in the LVD detector \cite{Aga07}.
 Lazauskas \etal~\cite{Laz07} used QRPA  with the Skyrme force
to explore the possibility of
performing nuclear structure studies using neutrinos from low
energy beta-beams ~\cite{Vol04}.
 Several $\nu$-nucleus cross sections
for different nuclei were also obtained recently with
the relativistic QRPA (RQRPA)~\cite{Paa07}.
The $\nu_e-^{56}$Fe cross
sections were also described with the gross theory of
beta decay (GTBD)~\cite{Ito78}.

The cross section
for $\nu_e + (Z, A)\go (Z+1, A)+ e^-$ is given by
\br
\s(E_e,J_f)& = &\frac{|\pb_e| E_e}{2\pi} F(Z+1,E_e)
\int_{-1}^1
d(\cos\theta)\T_{\s}(|\kb|,J_f),\nn\\
\label{1}\er
where $F(Z+1,E_e)$ is the usual scattering Fermi
function, $k=p_e-q_\nu$ is the momentum transfer, $p_e$ and $q_\nu$
are the corresponding electron and neutrino momenta, and
$\theta\equiv \hat{\qb}_\nu\cdot\hat{\pb}_e$ is the angle between
the incident neutrino and emerging electron. The $\sigma(E_e, J_f)$
cross section is obtained within first-order perturbation theory
according to Ref.~\cite{Krm05}, where velocity-dependent terms are
included in the weak effective Hamiltonian. The transition amplitude
$\T_{\s}(|\kb|,J_f)$ depends on the neutrino leptonic traces and on
the nuclear matrix elements (NME), as explained in
Ref.~\cite{Krm05}. Here, the NME are evaluated in QRPA and in PQRPA.
We employ the $\delta$-interaction (in MeV fm$^3$)
\[
V=-4 \pi \left(v_sP_s+v_tP_t\right) \delta(r),
\]
with  different coupling constants $v_s$ and $v_t$ for the
particle-hole, particle-particle, and pairing channels.
This interaction was used in Refs.~\cite{Hir90a,Hir90b,Krm92,Krm94}
leading to a good description of single and double $\beta$-decays.

For $^{56}$Fe we work within a configuration space of 12
single-particle levels, including the oscillator shells $2\hbar\w$,
$3\hbar\w$ and $4\hbar\w$. The single-particle energies of the
active $3\hbar\w$  shell correspond to the experimental energies of
$^{56}$Ni. For the other $2\hbar\w$ and $4\hbar\w$ shells we have
used the harmonic oscillator energies with $\hbar\w/{\rm
MeV}=~45~A^{1/3}-25~A^{2/3}$. The parameters $v_s^{pair}(p)$ and
$v_s^{pair}(n)$ were obtained with the procedure of
Ref.~\cite{Hir90}, \ie by fitting the experimental gap pairing
energies of protons and neutrons, $\Delta_{n,p}(N, Z)$ (eq.(2.96) of
Ref.~\cite{Boh69}), to $\Delta_{n,p}$ defined by the usual BCS
equations. The BCS or PBCS equations were solved in the full space
of three oscillator shells. For the particle-hole channel we have
used $v_s^{ph}=27$ and $v_t^{ph}=64$ (in MeV fm$^3$). These values
were fitted to
$^{48}$Ca from a systematic study of the GT resonances~\cite{Krm94}
and shown to yield a good description of double beta decay. For the
particle-particle channel, it is convenient to define the parameters
\[
s=\frac{2v_s^{pp}}{v_s^{pair}(p)+v_s^{pair}(n)},
\]
and
\[t=\frac{2v_t^{pp}}{v_s^{pair}(p)+v_s^{pair}(n)},\]
associated to the coupling constant of the $T=1, S=0$
(singlet) and $T=0, S=1$ (triplet) channels respectively Ref.~\cite{Krm94}.
We adopt $s\simeq 1$, which restores the isospin symmetry in QRPA for $N>Z$.
As the experimental errors in the averaged cross sections are very
large, the agreement of the theoretical cross section is not
sufficient to select the best nuclear structure calculation and
other observables must be found. We use the behavior of the
$B(GT^-)$ strength as function of the parameter $t$ to conclude that
better results could be obtained when the particle-particle channel
is off, $t=0$. With this value the theoretical value($B(GT^-)=
17.7$) overestimates the experimental value ($9.9\pm2.4$
~\cite{Rap83}) similarly to previous and more sophisticated QRPA
calculations for $^{56}$Fe ($B(GT^-)= 18.68$ ~\cite{Sar03}) with the
Skyrme force.

The flux averaged inclusive cross section reads
\be
\langle \s_e  \rangle= \int dE_{\nu}
\s(E_\nu) n(E_{\nu}),
\label{2}\ee
where
$
\s_e(E_{\nu})=\sum_{J^{\pi}_f}
\s_e(E_e,J_{f}^\pi),
$
is the inclusive cross section as a function of the neutrino energy and
$n(E_{\nu})$ is the neutrino normalized flux. As a first test,
we fold the $\s_e(E_\nu)$ with the Michel energy
spectrum~\cite{Kol99a},
\be
n(E_\nu)=\frac{96 E_\nu^2}{M_\mu^4} \left(M_\mu-2 E_\nu\right),
\label{3}\ee
where $M_\mu$ is the muon mass.
In Table~\ref{tab:2} we compare our
$^{56}$Fe$(\nu_e,e^-)^{56}$Co cross section $\langle\s_e\rangle$
in QRPA and PQRPA with other nuclear models
for the energy window
of $\mu$-Decay-At-Rest (DAR) neutrinos that the KARMEN experiment observed.

\begin{figure}[t]
\vspace{-2.cm}
\begin{center}
{\includegraphics[width=8cm,height=9.cm]{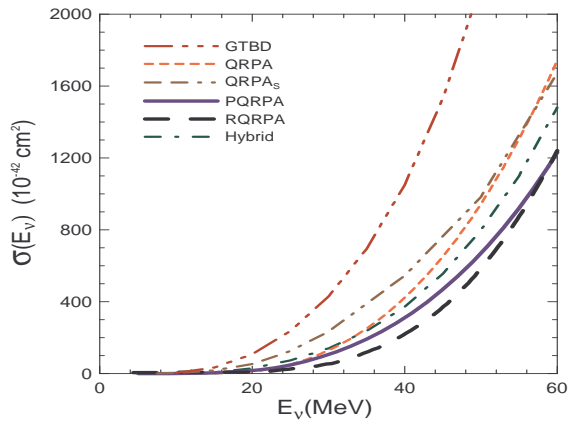}}
\end{center}
\vspace{-2.cm}
\caption{\label{fig1}(Color online) Inclusive
$^{56}$Fe$(\nu_e,e^-)^{56}$Co cross sections
(in $10^{-42}$ cm$^2$) evaluated in QRPA (dashed line)
and PQRPA (solid line), in the DAR region, are
compared with those obtained with other nuclear structure
models: GTBD (dashed-dot-dot-dot line)~\cite{Ito78},
Hybrid (dashed-dot line)~\cite{Kol01},
QRPA$_{\sss S}$ (dashed-dot dot line)~\cite{Laz07}, and
RQRPA(dashed line)~\cite{Paa07}.}
\end{figure}

\begin{table}[h]
\begin{center}
\caption{\label{tab:2} Comparison of
$\langle\s_e\rangle$ in $10^{-42}$~cm$^{2}$ for
$^{56}$Fe$(\nu_e,e^-)^{56}$Co
obtained in our QRPA and PQRPA confronted
with other nuclear models. For the Hybrid model~\cite{Kol99a}
we denote with (a)  partial occupation
in the ground state and with (b) no occupation.
}
\newcommand{\cc}[1]{\multicolumn{1}{c}{#1}}
\renewcommand{\tabcolsep}{2.6 pc} 
\renewcommand{\arraystretch}{1.1} 
\bigskip
\begin{tabular}{l c } \hline
Model&$\langle\s_e\rangle$
\\ \hline
QRPA&$264.6$
\\
PQRPA&$197.3$
\\
Hybrid$^{(a)}$~\cite{Kol99a}&$228.9$
\\
Hybrid$^{(b)}$~\cite{Kol99a}&$238.1$
\\
TM~\cite{Min02}&$214$
\\
RPA~\cite{Ath06}&$277$
\\
QRPA$_{\sss S}$~\cite{Laz07}&$352$
\\
RQRPA~\cite{Paa07}&$140$
\\
Exp\cite{Mas99}&$256\pm108\pm43$
\\ \hline
\end{tabular}\end{center}\end{table}
From Table \ref{tab:2} we note that
our results for $\langle\s_e\rangle= 264.6 \x 10^{-42}$ cm$^2$ (QRPA),
and $197.3 \x 10^{-42}$ cm$^2$ (PQRPA) are in
agreement with the experimental value
$256 \pm 83(stat) \pm 42(syst) \x 10^{-42}$ cm$^2$ \cite{Mas99}.
The main difference between QRPA and PQRPA, both solved
consistently with the same interaction, shows that
the projection procedure is important in a medium mass nucleus such
as $^{56}$Fe.

In Figure~\ref{fig1} we plot the inclusive
$^{56}$Fe$(\nu_e,e^-)^{56}$Co cross sections
(in $10^{-42}$ cm$^2$) evaluated in QRPA (dashed line)
and PQRPA (solid line), in the DAR region.
A comparison is shown with other nuclear structure
models.
 All models yield the same energy dependence that goes
approximately as $E_\nu^2$  for low incident neutrino energies,
except for the GTBD model, which shows a large deviation
from the other cross sections.

\begin{figure}[t]
\vspace{-2.cm}
\begin{center}
{\includegraphics[width=8cm,height=9.cm]{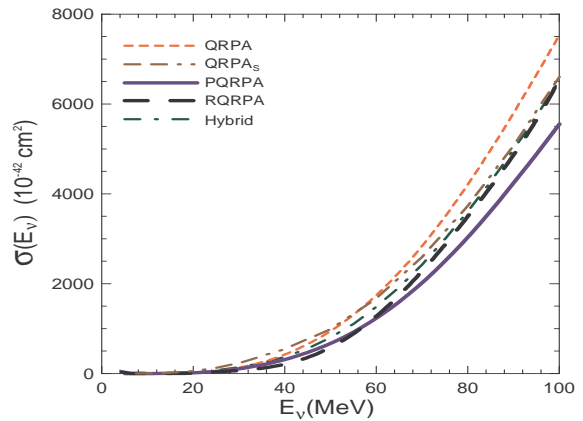}}
\end{center}
\vspace{-2.cm}
\caption{\label{fig2}(Color online)
Same as in Figure~\ref{fig1}. The
inclusive cross sections
(in $10^{-42}$ cm$^2$) are shown.
In this case, the neutrino energy window is
characteristic of the LVD experiment.}
\end{figure}

Figure \ref{fig2} excludes the GTBD results and extends the
energy scale to $100$ MeV (supernova neutrino energies).
$\s(E_\nu)$ reaches the DAR energy region at
$E_\nu \sim 60$ MeV. Beyond that the QRPA result is above
the other models. Nevertheless, the PQRPA cross section lies
below, showing the effect of the projection procedure.
In this region, the main
contribution arises from the non-allowed transitions, as found in previous
works~\cite{Laz07,Paa07}.

The number of events detected for supernova explosions is calculated as,
\be
N_{\a}=N_t \int_0^\infty  {\cal F}_\a(E_\nu) \cdot \sigma(E_\nu)
\cdot \epsilon(E_\nu) dE_\nu,
\label{4}\ee
where the index $\a=\nu_e,{\bar \nu}_e,\nu_x$ and
$(\nu_x=\nu_\tau,\nu_\mu,{\bar \nu}_\mu,{\bar \nu}_\tau)$
indicates the neutrino or antineutrino type,
$N_t$ is the number of target nuclei, ${\cal F}_\a(E_\nu)$ is the
neutrino flux, $\sigma(E_\nu)$ is the neutrino-nucleus cross section,
$\epsilon(E_\nu)$ is the detection efficiency, and $E_\nu$ is the neutrino energy.
Recent calculations by the LVD group
\cite{Aga07} estimate that the $(\nu_e+{\bar \nu}_e)$ interactions
on $^{56}$Fe are almost 17\% of the total detected signal.

The time-spectra can be approximated
by the zero-pinched Fermi-Dirac distribution.
For the neutrino of flavor $\a$, it is
\be
{\cal F}^0_\a(E_\nu, T_{\nu_\a})=\frac{L_\a}{4\pi D^2T^4_\a F_3(0)}
\frac{E_\nu^2}{\exp{(E_\nu/T_\a)}+1},
\label{5}\ee
where $D$ is the distance to the supernova, $E_\nu$ is the neutrino
energy, $L_\a$ is the time-integrated  energy of flavor $\nu_\a$,
$T_\a$ is the neutrino effective temperature, and
$F_3(0)\equiv  \int_0^\infty d^3x~ x^3/(e^{x}+1)$
is the normalization factor. For a galactic supernova explosion
at a typical distance $D=10$ kpc, it was assumed
that the total binding energy for each flavor
is $L_\a=f_{\nu_\a} E_b$, with
$E_b=3 \x 10^{53}$ erg, and a perfect energy equipartition between
the neutrino flavors, $f_\a=f_{\nu_e}=f_{{\bar \nu}_e}=f_{\nu_x}=1/6$.
Hence, it is possible to assume that the fluxes
$(\nu_\mu,\nu_\tau,{\bar \nu}_\mu,{\bar \nu}_\tau)$ are identical.
Because the pinched factor was assumed zero for all neutrino flavors,
we can fix the effective neutrino temperature as
$T_{\nu_x}=1.5 T_{{\bar \nu}_e}$ and $T_{\nu_e}=0.8 T_{{\bar \nu}_e}$,
leaving $T_{{\bar \nu}_e}$ as a variable parameter in the
interval studied in Ref.~\cite{Aga07}.

When the neutrinos escape from the star, they cross regions of
different densities where a flavor transition could happen. Usually
one assumes two resonance layers which we call
Mikheyev-Smirnov-Wolfenstein (MSW) resonances throughout the
text (see for example Ref.~\cite{Akh00}). According to the  mass
scheme shown in \cite{Aga07}, the observed electron neutrino fluxes
(${\cal F}_{\nu_e},{\cal F}_{{\bar \nu}_e}$) originating from MSW
resonances are linear combinations of the original neutrinos fluxes
in the star, ${\cal F}^0_{\nu_e}$ and ${\cal F}^0_{\nu_x}$, with
coefficients governed by the crossing probability in the high
density resonance layer, $P_{\sss H}(\Delta_{atm}^2,\theta_{13})$.
For simplicity, we only show differences that appear in the number
of events calculated from the convolution of cross sections obtained
with different nuclear structure models with the original supernova
fluxes, \ie
\be
N_{e}\equiv N_e(T_{\nu_e})=
N_t \int_0^\infty  {\cal F}^0_e(E_\nu, T_{\nu_e}) \cdot \sigma_e(E_\nu)
\cdot \epsilon(E_\nu) dE_\nu,
\label{6}\ee
for ``direct" electron neutrino event, and
\be
{\tilde N}_{e}\equiv {\tilde N}_{e}(T_{\nu_x})=
N_t \int_0^\infty  {\cal F}^0_x(E_\nu, T_{\nu_x}) \cdot \sigma_e(E_\nu)
\cdot \epsilon(E_\nu) dE_\nu,
\label{7}\ee
for the ``indirect"  number of events for electron neutrino
associated to the total $\nu_e$-flux coming from the
contribution of ${\cal F}^0_x$. Due to the MSW effect, electron
neutrino fluxes mix with non-electron neutrino fluxes
(\ie $\nu_x \equiv \nu_\mu, \nu_\tau$),
and therefore with the MSW resonance the $\nu_e$'s might get
a ``hot" contribution to their flux.
Another important issue, not considered for simplicity in
the present work, is the spectral swapping of the neutrino flux
(Ref.~\cite{Dun07}). Duan~\etal~ have shown that certain numerical
results in the simulation of neutrino and antineutrino flavor
evolution in the region above the post-supernova explosion
proto-neutron star cannot be easily explained with the conventional
MSW mechanism Ref.~\cite{Dun06}.

For the neutrino reactions $^{56}$Fe$(\nu_e,e^-)^{56}$Co, we
calculate $N_{e}$ and ${\tilde N}_{e}$ as a function of
the neutrino temperatures $T_{\nu_e}$ and $T_{\nu_x}$, folding $\sigma_e(E_\nu)$
from different nuclear structure models with the neutrino fluxes
${\cal F}^{0}_{\nu_e}(E_\nu,T_{\nu_e})$ and ${\cal F}^{0}_{\nu_x}(E_\nu,T_{\nu_x})$,
respectively. The limits for the temperatures, $T_{\nu_e}$ and $T_{\nu_x}$, were
obtained from the interval $T_{{\bar \nu}_e}\in[4,7]$ MeV
and the relations $T_{\nu_x}=1.5 T_{{\bar \nu}_e}$ and
$T_{\nu_e}=0.8 T_{{\bar \nu}_e}$, employed by the LVD group~\cite{Aga07}.
The $\e(E_\nu)$ efficiency is taken from Figure 1b of Ref.~\cite{Aga07}.
\begin{figure}[t]
\begin{center}
{\includegraphics[width=8cm,height=9.cm]{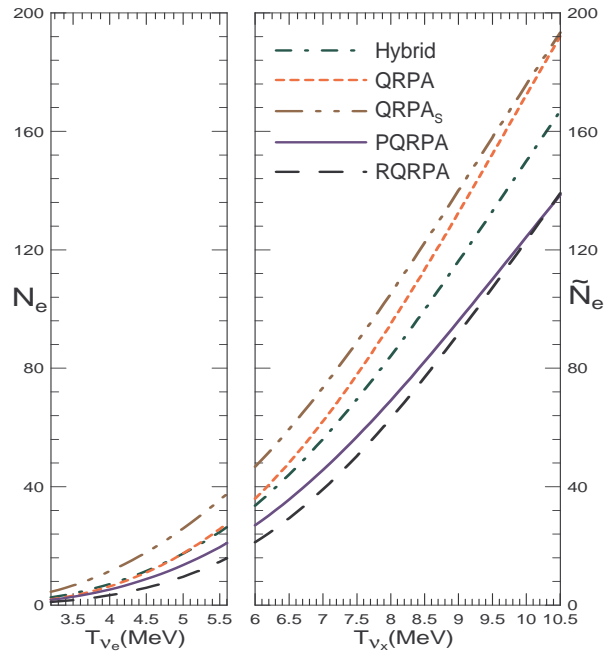}}
\end{center}
\vspace{-0.5cm}
\caption{\label{fig3}(Color online)
Number of events obtained from the convolution of the neutrino fluxes with
the cross section obtained with different nuclear structure
models: Hybrid (dashed-dot line)~\cite{Kol01},
QRPA (dashed line), QRPA$_{\sss S}$ (dashed-dot dot line)~\cite{Laz07},
PQRPA (solid line), and RQRPA(dashed line)~\cite{Paa07}.
}
\end{figure}
The results obtained are shown in Figure \ref{fig3}. The left panel shows
the number of events for electron neutrinos, $N_{e}$, with different
$\s_e(E_\nu)$, our QRPA and PQRPA,
QRPA$_{\sss S}$~\cite{Laz07}, RQRPA~\cite{Paa07} and the Hybrid
model~\cite{Kol99a} employed by the LVD detector. On the right panel
we show the number of events ${\tilde N}_{e}$. Although one
knows that $\nu_x$ neutrinos at supernova energies can only induce
neutral current events, we evaluate this quantity because it will be
modified by MSW oscillations according to the scheme presented in
equations (2) and (4) of Ref.~\cite{Aga07}, or in equations (10) and
(12) of Ref.~\cite{Fog05}.
Despite it is certain that the ${\tilde N}_e$ could be
obtained from the expression for $N_e$ extending the
interval of $T_{\nu_e}$ to cover the interval $T_{\nu_x}$,
this region ($T_{\nu_e} \in [6,10.5]$ MeV)  of
temperature for $\nu_e$ is not in agreement with the
physical range depending on the neutrino transport
that it is predicted by different
supernova modelers, as such in Ref.~\cite{Kei03}.

We note that $N_e$ and ${\tilde N}_e$ increase with
the temperatures $T_{\nu_e}$ and $T_{\nu_x}$.  The increase
for each $N_e$ follows the increase of the different $\sigma_e$.
The contribution of the neutrino
flux, ${\cal F}_{\nu_e}^0$, in  $N_e$ is strongly concentrated
in the region below $60$ MeV. This is because:
(i) the mean neutrino energy $\langle E_{\nu_e}\rangle$
of the flux varies from $10.1$ to $17.6$ MeV approximately
\footnote{The mean neutrino energy results
$\langle E_{\nu_e}\rangle \approx 3.15 T_{\nu_e}$ with
the pinching parameter $\eta=0$.};
and (ii) the contribution of the product of $\sigma_e$ with
the flux tail is not important.
Notice that the ordering of the $N_e$ in
Figure~\ref{fig3} is the same as the ordering of
$\sigma_e$ shown in Figure~\ref{fig1}.
For example, the crossing between
$N_e$'s of our QRPA and Hybrid model
at $T_{\nu_e}\sim 4.8$ MeV originates from
the crossing of the corresponding $\s_e$ at
$E_{\nu} \sim 32$ MeV as Figure~\ref{fig1} shows.

The above behavior also applies to ${\tilde N}_e$, but
they are shifted according to the shift that
${\cal F}_{\nu_x}^0$  has with respect to ${\cal F}_{\nu_e}^0$.
This means that the
main contribution to ${\tilde N}_e$ comes from the convolution of
${\cal F}_{\nu_x}^0$ with $\s_e$ in the energy interval $[18.9, 33.1]$~MeV
where the larger energy flux of $\nu_x$ is concentrated.
The right panel of Figure~\ref{fig3} shows additional crossings at
 $T_{\nu_x} \sim 10.5$ MeV which is a result of
the corresponding crossings  of $\s_e$(QRPA-QRPA$_{\sss S}$)
at $E_\nu \sim 56$ MeV and $\s_e$(PQRPA-RQRPA) at
$E_\nu \sim 60$ MeV, shown in Figure~\ref{fig1}. We conclude that
the relevant energy interval for the $\nu_e-^{56}$Fe reaction is
$E_\nu \leq 60$ MeV for the astrophysical parameters adopted in LVD.

In summary, we have employed the projected QRPA to calculate the
$^{56}$Fe$(\nu_e,e^-)^{56}$Co cross section. The calculated cross
section is compared with a QRPA calculation with the same
interaction showing that the projection procedure is important for
medium mass nuclei. The cross section is also compared with other
RPA and Hybrid models. The PQRPA yields smaller cross section than
almost all RPA models with exception of relativistic
QRPA~\cite{Paa07} for $E_\nu \leq 60$ MeV.  Above this energy and up
to $E_\nu = 100$ MeV, the PQRPA leads to the smallest cross section.
Therefore, we feel that a more detailed study of allowed and
forbidden transitions in the region below $E_\nu=~100$ MeV is
imperative, both experimentally and theoretically. In particular,
the region with $E_\nu \leq 60$ MeV is the most important for the
LVD detector~\cite{Aga07}. In a future work we plan to include the
MSW effect in the same way as was as done by Agafonova~\etal~ and an
explicit account of the uncertainties in the supernova neutrino flux
will be considered.

This work was partially supported by the U.S. DOE grants
DE-FG02-08ER41533 and DE-FC02-07ER41457 (UNEDF, SciDAC-2).

\end{document}